%% file: cpt.tex
\documentclass{iopart}
\expandafter\let\csname equation*\endcsname\relax 
\expandafter\let\csname endequation*\endcsname\relax 
\usepackage{amsfonts,amsmath}
\usepackage{graphicx}
\newcommand{\adb}{\allowdisplaybreaks }
\newcommand{\TE}{{\rm TE}}
\newcommand{\TM}{{\rm TM}}
\newcommand{\te}{\textsf{\scriptsize TE} }
\newcommand{\tm}{\textsf{\scriptsize TM} }
\begin{document}
\title{The thermal Casimir-Polder interaction of an atom with spherical plasma shell}
\author{Nail R. Khusnutdinov\footnote{e-mail: 7nail7@gmail.com}}

\address{Institute of Physics, Kazan Federal University, Kremlevskaya 18,
Kazan, 420008, Russia}
\ead{7nail7@gmail.com}

\begin{abstract}
The van der Waals and Casimir-Polder interaction energy of an atom with an infinitely thin sphere with finite conductivity is investigated in the framework of the hydrodynamic approach at finite temperature. This configuration models the real interaction of an atom with fullerene. The Lifshitz approach is used to find the free energy. We find the explicit expression for the free energy and perform the analysis of it for i) high and low temperatures, ii) large radii of sphere and ii) short separation between an atom and sphere. At low temperatures the thermal part of the free energy approaches zero as forth power of the temperature while for high temperatures it is proportional to the first degree of the  temperature. The entropy of this system is positive for small radii of sphere and it becomes negative at low temperatures and for large radii of the sphere.  

\end{abstract}  





\section{Introduction}

Van der Waals dispersion forces plays an important role in different physical,   biological as well as chemical phenomena \cite{Milonni:1994:QVItQE,Parsegian:2006:VdWFHBCEP,Bordag:2009:ACE,Klimchitskaya:2009:TCfbrmet}. These forces arise between objects due to correlated quantum fluctuations. They depend on the shape of the body and their structure. The last achievements in Casimir effect and Casimir-Polder interaction have been discussed in great depth in books and reviews (see, for example recent book \cite{Bordag:2009:ACE}). Of special interest is dispersion forces between microparticles and macroscopic bodies. In the case of interaction between particle and plate it is commonly referred to as the Casimir-Polder force  \cite{Casimir:1948:TIrLdWf}. At short range the energy rises as third power of inverse distance between microparticle and the plate. The retardation of the interaction should be taken into account at large distances and the interaction energy falls down as the fourth power of distance. At separations larger than a few micrometers thermal effects become dominating. 

Cavity quantum electrodynamic effects inside a spherical cavity has been considered in Refs. \cite{Barut:1987:QebsSec,Jhe:1996:Cqeihsc,Dung:2000:Sdpdabtatsc}. The non-retarded van der Waals potential for an atom inside and outside a metallic bubble was considered in Ref. \cite{Boustimi:2000:Saioodomb}. The corrections for proximity force  approximation of the Casimir interactions of an object inside and outside a spherical metal shell were calculated in Ref. \cite{Zaheer:2010:Cioisms,Zaheer:2010:Cpcoebsc}. The perfectly conductive sphere  with an atom was investigated in Ref. \cite{Taddei:2010:itdi}. In Refs.  \cite{Sambale:2010:Cibasms,Ellingsen:2012:Cesopnm} the Casimir-Polder interaction between an atom and a magnetodielectric sphere was considered on the basis of point-scattering techniques. The consideration was related to results obtained in Ref. \cite{Messina:2009:Dibans} for interaction an atom with curved surface and Refs. \cite{Buhmann:2004:vdWeasnadb,Buhmann:2006:vdWeasnadb-err} for an atom with dielectric sphere. The Casimir and Casimir-Polder energy were considered in the Refs. \cite{Bordag:2008:Ovesps,Khusnutdinov:2011:vdWibasps} in the framework of the plasma (hydrodynamic) model of the sphere which was developed in the Barton's papers       
\cite{Barton:2004:Cesps,Barton:2005:CefpsIE} in the context of the fullerene molecule $C_{60}$. The Barton's approach was based on the Fetter paper \cite{Fetter:1973:ElegISl} which was devoted to the properties of the two-dimensional free-electron gas. This model was exploited in Ref. \cite{Bordag:2005:Safpsm} for flat sheet. Complete list of references about particle in the spherical geometry may be found in reviews \cite{Buhmann:2007:Dfmqe,Scheel:2008:MQECA}.

The spherical plasma shell was considered in the framework of this model in Ref. \cite{Bordag:2008:Ovesps} in context of the Casimir energy and in Ref. \cite{Khusnutdinov:2011:vdWibasps} for the Casimir-Polder interaction of an atom with spherical plasma surface. In the context of this model the conductive sphere is represented by the two-dimensional surface with free-electron gas on this surface. The dynamic of this gas is described by hydrodynamics. All information about electron gas is encoded solely by parameter with dimension of wave number: $\Omega = 4 \pi n e^2/m c^2$, where $n$ is surface density of electron gas, $e$ and $m$ are the respectively charge and mass of electron. It was shown that the interaction energy between an atom and the sphere has the form of the Casimir-Polder energy for an atom and a plate if atom is situated close to the sphere. Away from the sphere the energy has the form of the atom-atom interaction energy with an expression for the polarizability of the sphere. The numerical simulations were made for the sphere with parameters of the fullerene molecule $C_{60}$ and hydrogen atom.  

%

In the present paper we consider the thermal Casimir-Polder interaction energy for a system comprising of an atom and conductive sphere. The sphere is described by the plasma (hydrodynamical) model which was considered in detail in the Refs. \cite{Barton:2004:Cesps,Barton:2005:CefpsIE} and we use the Lifshitz approach  \cite{Lifshitz:1956:Ttmafbs,Lifshitz:1980:SP} to calculate the free energy of this system. We outline the Lifshitz approach in the Sec. \ref{sec:2}. 

The calculations performed in the paper could find a potential application to interaction microparticles with molecule of fullerene. The Casimir-Polder interaction of the fullerene molecules $C_{60}$ and $C_{70}$ with gold and silicon nitride surfaces was considered in detail in Ref. \cite{Buhmann:2012:Cifms}. The fullerene molecule is graphene shaped as a sphere and the more realistic model to calculate the Casimir-Polder interaction between an atom and a fullerene is the Dirac model. In this model the frequency spectrum has linear form and it is valid for low energies. Some calculations related with the Casimir and Casimir-Polder energies were made in the framework of this model in the Refs.  \cite{Bordag:2009:CibpcgdbDm,Fialkovsky:2011:FCeg,Churkin:2010:ChDmdibgHtoNa}. The frequency spectrum in the hydrodynamic (plasma) model has quadratic form and the model is high-frequency approximation of the theory. Both models are in qualitative agreement with respect to the Casimir-Polder interaction. The calculations performed in the Ref.  \cite{Churkin:2010:ChDmdibgHtoNa} showed that the Casimir-Polder force calculated in framework the Dirac model is smaller then for hydrodynamic model.        

The paper is organized as follows. In Sec. \ref{sec:2} we discuss the Lifshitz approach and derive the expression for the free energy and represent it in different forms.
Section \ref{sec:3} is devoted to consideration of the specific limit cases. We consider the ideal case with infinite radius of the sphere but finite distance between the surface and  the atom and obtain the first correction term over inverse radius of the sphere. We obtain also the low and high temperature expansions and consider the case of the short distance from the sphere. The expression for the entropy is found in Sec. \ref{sec:4} and it is analyzed in the limits of small and high temperatures. We show plots of the numerical calculations for the system comprised the hydrogen atom and sphere with parameters of the  fullerene $C_{60}$. In Sec. \ref{sec:5} we discuss the results obtained.

\section{Lifshitz approach for an atom near the plasma spherical shell}
\label{sec:2}
\begin{figure}[ht]
\centerline{\includegraphics[width=7cm]{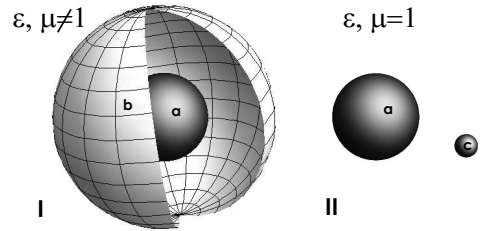}}
\caption{We adopt the Lifshitz approach. I: \textit{a} - the plasma sphere of radius $R$, \textit{b} - a spherical cavity of radius $L = R+d$ inside the infinite  material with $\varepsilon, \mu \not = 1$. We rarefy the media $\varepsilon =  1 + 4\pi N \alpha + O(N^2)$ and obtain configuration II: plasma sphere with single atom \textit{c} at the same distance $d$ from the sphere as was cavity's surface.} \label{fig:lifshits}
\end{figure}

We adopt here the approach developed by Lifshitz in Refs. \cite{Lifshitz:1956:Ttmafbs,Lifshitz:1980:SP}. This approach was applied to calculation the Casimir-Polder interaction energy of an atom and plasma sphere at zero temperature in Ref. \cite{Khusnutdinov:2011:vdWibasps}. Let us shortly discuss the approach. We consider a conductive infinitely thin sphere with radius $R$ inside the vacuum spherical cavity with radius $L=R+d$ which is into the dielectric media with parameters $\mu, \varepsilon$ (see Fig. \ref{fig:lifshits}). Due to the spherical symmetry of the problem under consideration the electromagnetic field may be divided on the two polarizations which usually called as \TE\ and \TM\ modes. We have two concentric spheres and we should consider the boundary conditions on the two spherical boundaries. The full information about the problem is encoded in these boundary conditions.  
The information about plasma sphere is encoded in single parameter $\Omega = 4\pi n e^2/mc^2$ with dimension of the wave number, where $e$ is charge of the electron, $m$ -- its mass and $n$ is the surface density of electrons on the sphere. The applications of this model for vacuum fluctuations of field see in Refs. 
\cite{Barton:2004:Cesps,Barton:2005:CefpsIE,Bordag:2005:Safpsm,%
Bordag:2008:Ovesps}. 

The spectrum of the electromagnetic oscillations, $\omega = kc$, in this configuration may be found from these boundary conditions (see Ref. \cite{Khusnutdinov:2011:vdWibasps} for more details):
\numparts \label{eq:bound_cond}
\begin{eqnarray}
 \Sigma_{\te} &=& -i\left\{ H'(z_\varepsilon) \Psi_{\te} -
\frac{1}{\sqrt{\varepsilon}} H(z_\varepsilon) \Psi'_{\te}\right\} = 0 ,\adb\\
 \Sigma_{\tm} &=& -iz^2\left\{H(z_\varepsilon) \Psi'_{\tm} -
\frac{1}{\sqrt{\varepsilon}} H'(z_\varepsilon) \Psi_{\tm}\right\} = 0,
\end{eqnarray}
where 
\begin{eqnarray}
 \Psi_{\te}(z) &=& J(z) + \frac{Q}{x} J(x) [J(x) Y(z) - J(z) Y(x)],\adb\\
 \Psi_{\tm}(z) &=& J(z) + \frac{Q}{x} J'(x) [J'(x) Y(z) - J(z) Y'(x)],
\end{eqnarray}
\endnumparts 
and $J(x) = xj_l(x),\ Y(x) = xy_l(x),\ H(x) = xh^{(1)}_l(x)$ are the Riccati-Bessel functions. Here $x=kR,\ Q=\Omega R$ and $z_\varepsilon = z \sqrt{\varepsilon}$, $z = kL = \omega L/c$.     

The free energy, ${\cal F}$, of the system is expressed in terms of these frequencies  in the following form \cite{Bordag:2009:ACE}:
\begin{equation}
{\cal F} = k_B T \sum_{l=1}^\infty (2l+1)\sum_{n=1}^\infty \left[ \ln \left( 2 \sinh\frac{\hbar \omega_{n,l}^{\tm}}{2k_B T} \right) + \ln \left( 2 \sinh\frac{\hbar \omega_{n,l}^{\te}}{2k_B T} \right) \right],
\end{equation}
where index $n$ numbers the solutions of the boundary condition equations (\ref{eq:bound_cond}). Then we convert the sum over $n$ to the contour integral
\begin{equation}
{\cal F} = k_B T \sum_{l=1}^\infty (2l+1)\frac{1}{2\pi i}\oint_\gamma \ln \left( 2 \sinh\frac{\hbar \omega}{2k_B T} \right) d\left(\ln\Sigma_{\tm} + \ln \Sigma_{\te}  \right),
\end{equation}
where the contour $\gamma$ counter-clockwise encloses all real positive solutions of the Eq. (\ref{eq:bound_cond}). To put the contour to the imaginary axes we should clear up the zero frequency behaviour of functions $\Sigma$. If  $Q\not=0$ we obtain ($\nu = l + 1/2$)
\numparts 
\begin{eqnarray}
\fl \Sigma_{\te}|_{\omega\to 0} = \left(1+ \frac{Q}{2\nu}\right)\varepsilon^{-\frac{l+1}{2}}, \\
\fl \Sigma_{\tm}|_{\omega\to 0} = Q \frac{l(l+1) [(1+l)(1-\varepsilon^2) + a^{1+2l}(l+\varepsilon^2 (l+1))]}{4\nu^2} \chi^{1-2l} \varepsilon^{-\frac{l+2}{2}},
\end{eqnarray}
\endnumparts 
where $\chi=L/R = 1+d/R$, $Q=\Omega R$ and $\varepsilon$ is dielectric permittivity. In the case $Q=0$ we have 
\numparts 
\begin{eqnarray}
\Sigma_{\te}|_{\omega\to 0} &=& \varepsilon^{-\frac{l+1}{2}}, \\
\Sigma_{\tm}|_{\omega\to 0} &=& -\frac{l+\varepsilon^2(l+1)}{2\nu} \omega^2 \chi^2\varepsilon^{-\frac{l+2}{2}}.
\end{eqnarray}
\endnumparts 
Therefore, no poles appear at the point $\omega = 0$ and we may shift the contour $\gamma$ to the imaginary axes.  

In the imaginary axes the function under logarithm has zeros at the Matsubara frequencies 
\begin{equation}
\omega_n = i \xi_n = i \frac{2\pi n k_B T}{\hbar},\ n=0,\pm 1,\pm 2 \ldots .
\end{equation}
We put the contour to the right of the imaginary axes,
\begin{equation}
{\cal F} = k_B T \sum_{l=1}^\infty (2l+1)\frac{1}{2\pi i}\int_{\epsilon + i\infty}^{\epsilon - i\infty} \ln \left( 2 \sinh\frac{\hbar \omega}{2k_B T} \right) d\left(\ln\Sigma_{\tm} + \ln \Sigma_{\te}  \right),
\end{equation}
and then take the limit $\epsilon \to 0$. Integrating by part we obtain
\begin{equation}
{\cal F} = -k_B T \sum_{l=1}^\infty (2l+1)\frac{1}{2\pi i}\int_{\epsilon + i\infty}^{\epsilon - i\infty} \frac{\hbar}{2k_B T}  \coth\frac{\hbar \omega}{2k_B T}\left(\ln\Sigma_{\tm} + \ln \Sigma_{\te}  \right) d\omega.
\end{equation} 
Then we shift the contour $\gamma$ to the imaginary axes. We obtain infinite sum of integrals in the imaginary axes between the Matsubara frequencies and integrals over semi-spheres around these frequencies. The integrals between $\xi_n$ and $\xi_{n+1}$ and between $\xi_{-n}$ and $\xi_{-n-1}$ cancel each other and we obtain ($\nu = l+ 1/2$)
\begin{equation}
{\cal F} = k_B T \sum_{l=1}^\infty \nu \sum_{n=-\infty}^{+\infty} \left(\ln\Sigma_{\tm} + \ln \Sigma_{\te}  \right)_{\omega = i\xi_{|n|}}.
\end{equation} 

To calculate the free energy per unit atom, ${\cal F}^a$, let us rarefy the media with $\varepsilon (i\omega) = 1 + 4\pi N \alpha (i\omega) + O(N^2)$, where $\alpha$ is polarizability of the atom and the density of the dielectric matter $N\to 0$ (see Fig. \ref{fig:lifshits}). In this case the free energy, ${\cal F}$, is expressed in terms the energy per unit atom ${\cal F}^a$ by relation
\begin{equation}
{\cal F} = N \int_d^\infty {\cal F}^a 4\pi (R+r)^2 dr + O(N^2).
\end{equation} 
From this expression it follows that  
\begin{equation}
{\cal F}^a = - \lim_{N\to 0} \frac{\partial_d {\cal F}}{4\pi N (R+d)^2}.
\end{equation}
After straightforward calculations we obtain 
\begin{equation}
{\cal F}^a = \frac{k_B T}{(R+d)^2} \sum_{l=1}^\infty \nu \sum_{n=-\infty}^{+\infty} \left(\frac{k\alpha (i\omega)}{G_{TE}} + \frac{k\alpha (i\omega)}{G_{TM}}  \right)_{\omega = \xi_{|n|}},
\end{equation} 
where 
\begin{eqnarray*}
\fl G_{TE}^{-1} = e_l(z)s_l(z) - \frac{Q}{x} \frac{s_l^2(x)e_l^2(z)}{f_{TE}(ik)}, \\
\fl G_{TM}^{-1} = -e'_l(z)s'_l(z) - e_l(z)s_l(z) \frac{\nu^2 - \frac 14}{z^2} - \frac{Q}{x} \frac{{s'}_l^2(x){e'}_l^2(z) + {s'}_l^2(x)e_l^2(z) \frac{\nu^2 - \frac 14}{z^2}}{f_{TM}(ik)},
\end{eqnarray*}
and 
\begin{equation}
 s_l(x) = \sqrt{\frac{\pi x}{2}} I_{l+1/2}(x), \ e_l(x) = \sqrt{\frac{2 x}{\pi}}
K_{l+1/2}(x)
\end{equation}
are the modified Riccatti-Bessel spherical functions.

For renormalization we throw away terms which survive in the limit $Q\to 0$ and arrive with formula
\begin{eqnarray}\label{eq:Fgen}
{\cal F}^a &=& -\frac{2 k_B T\Omega}{(R+d)^2} \sum_{l=1}^\infty \nu \sum_{n=0}^{+\infty}{}' \alpha (i\xi_{n})\nonumber\\
&\times& \left\{ \frac{s_l^2(x) e_l^2(z)}{f_{TE}(ix)} + \frac{{s'_l}^2(x) {e'_l}^2(z) + {s'_l}^2(x) e_l^2(z) \frac{\nu^2 - \frac 14}{z^2}}{f_{TM}(ix)}\right\}_{\omega = \xi_{n}}, 
\end{eqnarray}
where $x=\omega R/c$ and $z=\omega L/c = \omega (R+d)/c$, 
\begin{eqnarray*}
f_{TE}(ix) &=& 1 + \frac{Q}{x} s_l(x) e_l(x),\\
f_{TM}(ix) &=& 1 - \frac{Q}{x}{s'_l}(x) {e'_l}(x),
\end{eqnarray*}
are the Jost functions on the imaginary axes, and the sum with prime means that zero term has factor $1/2$. The formula (\ref{eq:Fgen}) has the same form as for zero temperature case  obtained in Ref. \cite{Khusnutdinov:2011:vdWibasps} but with summation over Matsubara frequencies instead of integration.    

Now we use single-oscillatory approximation for $\alpha$:
\begin{equation}
\alpha (i\omega) = \frac{g_a^2}{\omega^2 + \omega_a^2}.
\end{equation}
At Matsubara frequencies we have
\begin{equation}
\alpha (i\xi_n) = \frac{g_a^2}{\omega_a^2} \frac{1}{\frac{\xi_n^2}{\omega_a^2} +1} = \frac{\alpha(0)}{\frac{\xi_n^2}{\omega_a^2} +1}.
\end{equation}
In the problem under consideration the three different effective temperatures appear
\begin{equation}
T_\omega = \frac{\hbar \omega_a}{2\pi k_B},\ T_R = \frac{\hbar c}{2\pi k_B R}, \ T_d = \frac{\hbar c}{2\pi k_B d}.
\end{equation}
The variables are expressed in the terms of these temperatures by relations
\numparts 
\begin{eqnarray}
\frac{\xi_n}{\omega_a} &=& \frac{2\pi k_B T n}{\hbar \omega_a} = n\frac{T}{T_\omega}, \\
x &=& \frac{\xi_n R}{c} = \frac{2\pi k_B T R n}{\hbar c} = n\frac{T}{T_R}, \\
z &=& \frac{\xi_n L}{c} = \frac{2\pi k_B T L n}{\hbar c} = n\left(\frac{T}{T_R} +  \frac{T}{T_d} \right).
\end{eqnarray}  
\endnumparts 
Let us estimate these temperatures for molecule $C_{60}$ and hydrogen atom. For molecule $C_{60}$ \cite{Barton:2004:Cesps} we have $R = 0.342 nm$ and the frequency $\omega_a$ for hydrogen atom reads \cite{Bordag:2006:LfgscnvdWCi,Blagov:2007:vdWibmscn,Rauber:1982:Sdibaam}  $\omega_a = 11.65 eV = 17.698 \cdot 10^{15} Hz$. Taking into account these parameters we obtain 
$T_\omega = 2.15\cdot 10^4 K,\ T_R = 1.06\cdot 10^6 K$. 

Now we obtain another representation of the formula (\ref{eq:Fgen}). We rewrite Eq. (\ref{eq:Fgen}) in the following form
\numparts \label{eq:Fa}
\begin{equation}
{\cal F}^a = -\frac{2k_B T Q\alpha (0) a^2}{R^3(1+r)^2}  \sum_{l=1}^\infty \nu \sum_{n=0}^{+\infty}{}' F_l(n),
\end{equation}
where
\begin{eqnarray}
F_l(n) &=& \frac{g_l(n)}{n^2 + a^2}, \\
g_l(n) &=& \left\{ \frac{s_l^2(x) e_l^2(z)}{f_{TE}(ix)} + \frac{{s'_l}^2(x) {e'_l}^2(z) + {s'_l}^2(x) e_l^2(z) \frac{\nu^2 - \frac 14}{z^2}}{f_{TM}(ix)}\right\}_{\omega = \xi_{n}}.  
\end{eqnarray}
\endnumparts 
Here $a = T_\omega/T$ and $r=d/R$. The function $g_l$ consists of two parts which comes from \TE\ and \TM\ polarizations of electromagnetic field, correspondingly. Hereafter we will denote these contributions as $g_l^\te$ and $g_l^\tm$.   

The function $g_l(n)$ has no poles in the imaginary axes. Taking into account the Abel-Plana formula we obtain
\begin{eqnarray} 
\fl \sum_{n=0}^{+\infty}{}' F_l(n) = \int_0^\infty  \frac{g_l(t)dt}{t^2 + a^2} + \frac{\pi}{2a} \frac{g_l(ia)+g_l(-ia)}{e^{2\pi a} - 1}\nonumber \\
-\frac{i}{2a}\int_0^\infty \left(\frac{g_l(it)-g_l(-it)}{e^{2\pi t} - 1}\right)'_t \ln \left|\frac{a+t}{a-t}\right| dt. 
\end{eqnarray}
In the first term we change integrand variable $t = a k/(c\omega_a)$ and the contribution of this term is exactly the same as was obtained in Ref. \cite{Khusnutdinov:2011:vdWibasps} for zero temperature 
\begin{equation}\label{eq:T0}
E_0 =  - \frac{\hbar c}{\pi R^3 (1+r)^2} \int_0^\infty dk \alpha (i\omega) e_0(k),
\end{equation}  
where 
\begin{equation}
e_0(t) = Q\sum_{l=1}^\infty \nu g_l(t).
\end{equation}
The second and third terms give the temperature corrections
\begin{eqnarray}
{\cal F}^T_1 &=& - \frac{\alpha(0)e_1(a)}{R^3(1+r)^2} \frac{\hbar \omega_a}{e^{2\pi a} -1},\nonumber \\ 
{\cal F}^T_2 &=& -\frac{\alpha(0)\hbar \omega_a}{\pi R^3(1+r)^2} \int_0^\infty \left(\frac{e_2(t)}{e^{2\pi t} - 1}\right)'_t \ln \left|\frac{a+t}{a-t}\right| dt\nonumber\\
&=& -\frac{\alpha(0)\hbar \omega_a}{\pi R^3(1+r)^2} \int_0^\infty \left(\frac{e_2(at)}{e^{2\pi at} - 1}\right)'_t \ln \left|\frac{1+t}{1-t}\right| dt, \label{eq:T2}
\end{eqnarray}
where
\begin{eqnarray*} 
e_1(t) &=& \frac{1}{2}Q\sum_{l=1}^\infty \nu \left[g_l(it) + g_l(-it)\right] = \Re e_0(it), \\
e_2(t) &=& \frac{-i}{2}Q\sum_{l=1}^\infty \nu \left[g_l(it) - g_l(-it)\right] = \Im e_0(it).
\end{eqnarray*}

We note that the functions $g_l(\pm ia)$ and $g_l(\pm iat)$ (also $e_1(a)$ and $e_2(at)$) have no dependence on the temperature. Indeed, at the beginning (\ref{eq:Fa}) they depend on $n$ via $x = n T/T_R$ and $z = n T/T_R (1+r)$, only. After applying the Abel-Plana formula we should replace $n\to \pm i a$ (or $n\to \pm i a t$). Therefore, we obtain that $x \to i q_a$ and $z \to iq_a (1+r)$ and whole dependence on the temperature is accumulated in the exponential factor the Eq. (\ref{eq:T2}) via the single parameter $a = T_\omega/T$.  

\section{Limit cases}\label{sec:3}
\subsection{An atom near the ideal plate: $\Omega\to \infty$ and $R \to \infty$}
First of all we have to compare our formulas with well-known case of an atom near the flat surface. There is a need to take the limit of infinite conductivity $\Omega\to \infty$ to obtain ideal surface case and then take the limit of infinite radius of the sphere, providing finite distance, $d$, between an atom and the surface. To take limits we use the expression for free energy in the form of Eq. (\ref{eq:Fgen}). Taking the limit  of infinite conductivity $\Omega\to \infty$ we obtain the following expression for the free energy
\begin{eqnarray}\label{eq:FgenB}
{\cal F}^a &=& -\frac{2 k_B T}{R^3\chi^2} \sum_{n=0}^{+\infty}{}' \alpha (i\xi_{n}) \sum_{l=1}^\infty \nu x \nonumber\\
&\times& \left\{ \frac{s_l^2(x) e_l^2(z)}{s_l(x) e_l(x)} - \frac{{s'_l}^2(x) {e'_l}^2(z) + {s'_l}^2(x) e_l^2(z) \frac{\nu^2 - \frac 14}{z^2}}{s'_l(x) e'_l(x)}\right\}_{\omega = \xi_{n}}, 
\end{eqnarray} 
where $x = \xi_n R/c, z = x\chi $ and $\chi = 1+r = 1+d/R$. 

To obtain expansion over $r=d/R \ll 1$ we use the uniform Debye expansions for modified Bessel functions from textbook \cite{Abramowitz:1970:eHMFFGMT}. Taking into account these expansions we obtain 
\begin{eqnarray}\label{eq:FgenB-Rinfthy}
{\cal F}^a = -\frac{2 k_B T}{R^3\chi^3} \sum_{n=0}^{+\infty}{}' \alpha (i\xi_{n}) \sum_{l=1}^\infty \frac{\nu^2}{t(z)} e^{-2\nu [\eta(z) - \eta(x)]} \nonumber \\
\times   \left\{ 1 + \frac{1}{12\nu} \left( 3[t(x)-t(z)] + [t^3(x)-t^3(z)] + 6 t^5(z)\right) + \ldots \right\}_{x\to \frac{x}{\nu}}, 
\end{eqnarray} 
where $t(x) = 1/\sqrt{1+x^2}$ and $\eta (x) = 1/t(x) + \ln\frac{x}{1+1/t(x)}$. Now we replace summation over $\nu = l+1/2$ to integration over $\nu$ and change the integrand variable $\nu$ to $k_\perp = \nu/R$. We observe that the uniform expansion over $1/\nu$ corresponds to the expansion over $1/R$. For this reason, in the second term of expansion we may set $z=x$ up to first power of $1/R$ and the expression in braces takes the form $1+t^5(\xi_n/ck_\perp)/2k_\perp R$. We expand the exponent in above formula over $d/R \ll 1$ and obtain
\begin{equation}
e^{-2\nu [\eta(z) - \eta(x)]} = e^{-2d k_\perp \sqrt{1 + \frac{\xi_n^2}{c^2k_\perp^2}}}
\left\{1 + \frac{d^2k_\perp}{R} t\left( \frac{\xi_n}{ck_\perp} \right) + \ldots\right\}.
\end{equation} 

Taking into account these expansions we obtain the following expansion the free energy up to $R^{-1}$: 
\begin{equation}\label{eq:FgenBR}
{\cal F}^a = -2 k_B T \sum_{n=0}^{+\infty}{}' \alpha (i\xi_{n}) \int_0^\infty d k_\perp k_\perp q_n e^{-2 d q_n} \left\{ 1 - 3 \frac{d}{R} + \frac{k_\perp^4}{2 R q_n^5} + \frac{d^2 k_\perp^2}{Rq_n}\right\}, 
\end{equation} 
where $q_n^2 = k_\perp^2 + \xi_n^2/c^2$. The first term of this expansion corresponds to the case of an atom near flat ideal surface and it has the same form as in Ref. \cite{Bezerra:2008:Ltaiatqr}. The next terms give corrections due to finite radius of the sphere.  

Let us discuss an error which appears when we replace summation by integration. The precision of calculations is down to $R^{-1}$. The second term of expansion over $1/\nu$ in Eq. (\ref{eq:FgenB-Rinfthy}) gives contribution $\sim R^{-1}$ and when we replace summation by integration the error will be outside the considered precision. Let us consider the first term  in Eq. (\ref{eq:FgenB-Rinfthy}):
\begin{equation}
\sum_{l=1}^\infty \nu \sqrt{\nu^2 + z^2} e^{-2\nu [\eta(z) - \eta(x)]}. 
\end{equation}
To estimate error we use the Abel-Plana formula:
\begin{equation}
\sum_{l=1}^\infty f(\nu) = -f(\frac{1}{2}) + \int_0^\infty f(\nu) d\nu -i \int_0^\infty \frac{f(it) - f(-it)}{e^{2\pi t} +1} dt 
\end{equation}
with $f(\nu) = \nu \sqrt{\nu^2 + z^2} e^{-2\nu [\eta(z) - \eta(x)]}$. It is easy to show that the term $f(\frac{1}{2})$ gives contribution $\sim R^{-2}$ which is outside of the  precision and we drop this term. The domain of the integration in the last term is to be divided for three intervals namely, $0< t <x$, $x < t < z=x\chi$ and $t >z$. Considering each interval separately we obtain 
\begin{equation}\label{eq:ints}
\int_0^x \frac{2 t \sqrt{z^2 -t^2}e^{-2f_1}}{e^{2\pi t} +1} dt + \int_x^z \frac{2 t \sqrt{z^2 -t^2}e^{-2f_2} \cos (2v_2)}{e^{2\pi t} +1} dt + \int_z^\infty \frac{2 t \sqrt{t^2 - z^2} \sin (2v_3)}{e^{2\pi t} +1} dt, 
\end{equation}
where the functions read
\begin{eqnarray*}
\fl f_1(t) = \sqrt{z^2 - t^2} - \sqrt{x^2 - t^2} - t \left(\arctan\frac{t}{\sqrt{x^2 - t^2}} - \arctan\frac{t}{\sqrt{z^2 - t^2}} \right),\\ 
f_2(t) = \sqrt{z^2 - t^2}- t \arctan\frac{\sqrt{z^2 - t^2}}{t},\\
v_2(t) = t\ln \frac{t + \sqrt{t^2 -x^2}}{x} - \sqrt{t^2 -x^2},\\
v_3(t) = \sqrt{t^2 -z^2} - \sqrt{t^2 -x^2} + t \ln\left[ \frac{z}{x} \frac{t + \sqrt{t^2 -x^2}}{t + \sqrt{t^2 -z^2}} \right].
\end{eqnarray*}
Now we take the limit $R\to \infty$ and expand above integrals over $1/R$. The expansion of the first integral reads: $x e^{-\tau}/24$, where $\tau = 4\pi k_B dT/\hbar c$ and $x = \xi_n R/c$. Because of the factor $R^{-3}$ in the free energy (\ref{eq:FgenB-Rinfthy}) the contribution of this integral is $\sim R^{-2}$. Both rest integrals in (\ref{eq:ints}) give exponentially small contributions $\sim e^{-2\pi x}$. Therefore, the error which appears when we replace summation to integration is order $R^{-2}$ and it is outside the precision under consideration.

\subsection{Low temperature expansion}

Let us consider now the low temperature expansion: $T\ll T_\omega, T_R$, and $T\ll T_{R+d}$. The last relation means that we consider distances $r = \frac{d}{R} \ll \frac{T_R}{T}$.  The first temperature correction given by Eq. (\ref{eq:T2}) gives exponentially small contribution, ${\cal F}^T_1 \sim e^{-2\pi T_\omega/T}$. To find expansion of the second temperature correction (\ref{eq:T2}) we take into account that the function $e_2(t)$ depends on the $t T/T_R$ and $t T/T_R (1+r)$. Owing to the exponential factor we have to take into account only small domain of integration $t\ll 1$ in the integral. Let us find the power expansion of the function $g_l(it)$ over the $t$ and keep the main terms. 

1. \TE\ contribution. The expansion begins with $s^5$:
\begin{equation}
e_2^\te (t) = -s^5 Q\frac{Q - 2 (3+Q)\chi^3}{6(3+Q)^2 \chi^2} + \ldots,
\end{equation}
where $s = tT/T_R$ and $\chi = 1 + r = 1+d/R$.

2. \TM\ contribution. The expansion begins with $s^3$:
\begin{equation}
e_2^\tm(t) = - s^3 \frac{2}{\chi^4} - s^5 \frac{15 Q + 72 (5+Q) \chi^2 + 40 Q \chi^4 + 88 Q \chi^7}{60 Q \chi^6} + \ldots.
\end{equation}
This expression holds for $T/T_R \ll Q$. The case $Q=0$ should be considered separately. We set $Q=0$ at the beginning and obtain   
\begin{eqnarray}
e_2^\tm |_{Q=0} &=& Q\left\{ s^3 \chi \frac{44}{45} - s^5 \chi \frac{107 + 184\chi^2}{315}  \right \}_{Q\to 0} = 0.
\end{eqnarray} 

\begin{figure}[ht]
\centerline{\includegraphics[width=8cm]{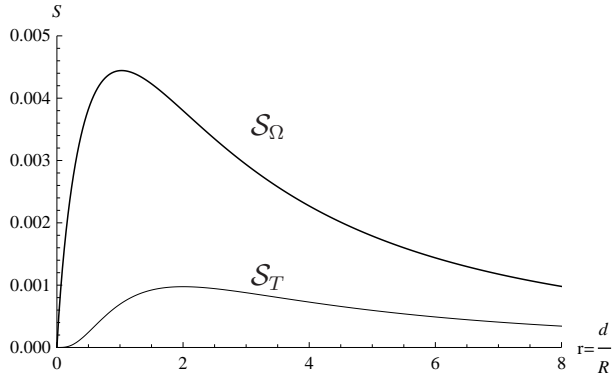}}
\caption{The plot of the functions ${\cal S}_\Omega$ and ${\cal S}_T$ from Eq. (\ref{eq:FaLowT}) calculated for molecule $C_{60}$ and hydrogen atom. The function ${\cal S}_\Omega$ is the zero temperature contribution and depends on the parameters of the sphere and the atom. The function ${\cal S}_T$ is universal and depends on distance $r$, only.}\label{fig:s}
\end{figure}

Therefore, the main contribution for the case of the finite conductivity, $Q>0$, comes from \TM\ mode and it reads
\begin{equation}
{\cal F}^T_2 = \frac{2\alpha(0)\hbar \omega_a}{\pi \chi^6 R^3} \frac{T^3}{T_R^3}\int_0^\infty \left(\frac{t^3}{e^{2\pi t} - 1}\right)'_t \ln \left|\frac{a+t}{a-t}\right| dt.
\end{equation}
Because of $a = T_\omega/T \gg 1$ we expand logarithm over $t\ll a$ and obtain
\begin{equation}
\int_0^\infty \left(\frac{t^3}{e^{2\pi t} - 1}\right)'_t \ln \left|\frac{a+t}{a-t}\right| dt = \int_0^\infty \left(\frac{t^3}{e^{2\pi t} - 1}\right)'_t \frac{2t}{a}  dt  = - \frac{1}{120 a}.
\end{equation}
Therefore, 
\begin{equation}\label{eq:FaLowT-A}
{\cal F}^T_2 = - \frac{\hbar c \alpha (0)}{60\pi R^4 (1+r)^6} \frac{T^4}{T_R^4} = - \frac{4\pi^3}{15}\frac{\alpha (0)}{(1+r)^6} \frac{(k_B T)^4}{(\hbar c)^3},
\end{equation}
and 
\begin{equation}\label{eq:FaLowT}
{\cal F}^a = E_{CP} \left\{{\cal S}_\Omega  + {\cal S}_T \left(\frac{T}{T_R} \right)^4 \right\},
\end{equation}
where 
\begin{equation}
{\cal S}_T  = \frac{2r^4}{45(1+r)^6},
\end{equation} 
and $E_{CP} = -3\hbar c \alpha (0)/8\pi d^4$ is the Casimir-Polder energy. The plots of ${\cal S}_\Omega$ and ${\cal S}_T$ are shown in the Fig. \ref{fig:s} for the case of the molecule $C_{60}$ and hydrogen atom. 

\subsection{High temperature expansion}

In this case it is better to use Eq. (\ref{eq:Fgen}):
\begin{eqnarray}
{\cal F}^a &=& -\frac{2 k_B T\Omega}{(R+d)^2} \sum_{l=1}^\infty \nu \sum_{n=0}^{+\infty}{}' \alpha (i\xi_{n})\nonumber\\
&\times& \left\{ \frac{s_l^2(x) e_l^2(z)}{f_{TE}(ix)} + \frac{{s'_l}^2(x) {e'_l}^2(z) + {s'_l}^2(x) e_l^2(z) \frac{\nu^2 - \frac 14}{z^2}}{f_{TM}(ix)}\right\}_{\omega = \xi_{n}}, 
\end{eqnarray} 
where $x=kR$ and $z=kL = k(R+d)$, and sum with prime means that zero term has factor $1/2$.  

At high temperatures the Matsubara frequencies  $\xi_n \sim n T \to \infty$ for $n>0$ and these terms in sum over $n$ give exponentially small contribution $\sim e^{-2\xi_nr}$. The main contribution comes from zero mode, $n=0$, and it maybe calculated in manifest form. Indeed, taking the limit $n\to 0$ in above expression we obtain the following contribution 
\begin{equation}\label{eq:FHighT}
{\cal F}^a = - \frac{k_B T \alpha (0)}{\chi^2 R^3} \sum_{l=1}^\infty  \frac{\nu(l+1)}{\chi^{2(l+1)}} =  - k_B T \alpha (0)\frac{6 r^4+24 r^3+33 r^2+18 r+4}{2 r^3 (r+1)^4 (r+2)^3 R^3},
\end{equation}
where $\chi = 1 + d/R$ and $r = d/R$. We have the following asymptotic for short and large separations between an atom and the sphere
\begin{eqnarray*}
{\cal F}^a &=& - \frac{k_B T \alpha (0)}{4 d^3},\ \ \frac{d}{R} \ll 1,\label{eq:FTsd} \\
{\cal F}^a &=& - \frac{3 k_B T \alpha (0) R^3}{ d^6} ,\ \ \frac{d}{R} \gg 1. 
\end{eqnarray*}

Extracting the Casimir-Polder energy we obtain
\begin{equation}
{\cal F}^a = E_{CP} \overline{{\cal S}}_T \frac{T}{T_R},
\end{equation}
where 
\begin{equation}
\overline{{\cal S}}_T = \frac{2r(6 r^4+24 r^3+33 r^2+18 r+4)}{3 (r+1)^4 (r+2)^3} ,  
\end{equation}

\subsection{Short distance behaviour, $r=\frac{d}{R} \ll 1$.} 

Let us consider Eq. (\ref{eq:T2}). In the limit $r\to 0$ the series is divergent for great $l$. For this reason we use standard formulas (9.3.1) from Ref.  \cite{Abramowitz:1970:eHMFFGMT}:
\numparts 
\begin{eqnarray}
J_\nu (z) &\approx & \frac{1}{\sqrt{2\pi \nu}} \left(\frac{ez}{2\nu} \right)^\nu,\\
Y_\nu (z) &\approx & -\frac{2}{\sqrt{2\pi \nu}} \left(\frac{ez}{2\nu} \right)^{-\nu},
\end{eqnarray} 
\endnumparts 
which are fulfil for $\nu\to \infty$. Taking into account these formulas in the expression for $g_l(ia)$ we obtain
\begin{equation*}
\nu \left\{g_l^\tm (ia) + g_l^\tm (-ia)\right\} \approx  \frac{8\nu^4 \chi^{-1-2\nu}}{(4\nu^2 -1)Q - 8\nu q_a^2}  \approx  \left\{ 
\begin{array}{ll}
+\frac{2\nu^2 \chi^{-1-2\nu}}{Q},& Q>0, r<Q  \\
-\frac{\nu^3 \chi^{-1-2\nu}}{q^2},& Q=0,r<q_a^2
\end{array} \right. 
\end{equation*}
for \TM\ contribution, and 
\begin{equation}
\nu \left\{g_l^\te (ia) + g_l^\te (-ia)\right\} \approx  -\frac{q_a^2 \chi^{1-2\nu}}{2\nu +Q} \approx  \left\{ 
\begin{array}{ll}
-\frac{q_a^2 \chi^{1-2\nu}}{2\nu},& Q\geq 0  \\
-\frac{q_a^2 \chi^{1-2\nu}}{Q},& Q\to \infty ,
\end{array} \right.  
\end{equation}
for \TE\ contribution. We make summation for finite $Q>0$ and $r<Q$ and obtain
\numparts  
\begin{eqnarray}
e_1^\tm (a)  &\approx &  \frac{1 - 2\chi^2 + 9 \chi^4}{4 \chi^2 (\chi^2 -1)^3} \approx \frac{1}{4r^3},\\
e_1^\te (a) &\approx &  \frac{Q q_a^2}{2} (1 - \chi \arctan\hspace{-.5ex} \textrm{h} \frac{1}{\chi}) \approx \frac{Q}{4}q_a^2 \ln r . 
\end{eqnarray}
For the case of the ideal surface, $Q\to \infty$, the \TM\ contribution will be the same, but contribution of \TE\ mode changes from logarithm to first inverse power    
\begin{equation}
e_1^\te (a) \approx   \frac{q_a^2}{4(1 - \chi^2)}  \approx -\frac{q_a^2}{8r}. 
\end{equation}
\endnumparts 

Therefore, for short distances $d \ll R$ and finite conductivity $Q>0$ (or $Q\to\infty$) the main contribution comes from \TM\ polarization and it reads 
\begin{equation}\label{eq:T1-1}
{\cal F}^T_1 \approx - \frac{\alpha(0)}{4d^3} \frac{\hbar \omega_a}{e^{\frac{\hbar \omega_a}{k_B T}} -1}.
\end{equation}
If $Q=0$ we have
\begin{equation}
e_1^\tm (a) \approx  \frac{1 - 5\chi^2 - 17 \chi^4 - 27 \chi^6}{8q_a^2 \chi^2 (\chi^2 -1)^4} \approx -\frac{3}{8q_a^2 r^4},  
\end{equation}
and therefore for $Q\to 0$ we obtain
\begin{equation}
{\cal F}^T_1 \approx \frac{3 Q \alpha(0)R}{16q_a^2d^4} \frac{\hbar \omega_a}{e^{\frac{\hbar \omega_a}{k_B T}} -1} \to 0
\end{equation}
as should be the case.

Let us consider Eq. (\ref{eq:T2}). By using the same asymptotic as above we observe that the series over $l$ in function $e_2(at)$ is convergent for any $r$ including $r=0$. Indeed, in the limit $l\to \infty$ we have
\begin{eqnarray}
-i \nu \left \{g_l^\te(iat) - g_l^\te(-iat)\right \} &\approx & \frac{q_a^2 \chi t^2}{2\nu} \left(\frac{etq_a}{2\nu} \right)^{2\nu}, \\  
-i \nu \left \{g_l^\tm(iat) - g_l^\tm(-iat) \right\} &\approx & -\frac{\nu^2 }{Q \chi} \left(\frac{etq_a}{2\nu \chi} \right)^{2\nu}.
\end{eqnarray}
Therefore the series is convergent as $l^{-l}$ even for $r=0\ (\chi =1)$ and this term gives finite contribution and we may threw away it comparing with terms considered above. 

The term with zero temperature contribution (\ref{eq:T0}) maybe considered in the same way. In the limit $r\to 0$ the series is divergent for large $l$. For this reason we use the Debye uniform expansion formulas from Ref.  \cite{Abramowitz:1970:eHMFFGMT} in which we change $z\to z/\nu$ and make the limit $\nu\to \infty$:
\numparts 
\begin{eqnarray}
I_\nu (z) &\approx & \frac{1}{\sqrt{2\pi \nu}} \left(\frac{ez}{2\nu} \right)^\nu,\\
K_\nu (z) &\approx & \frac{\pi}{\sqrt{2\pi \nu}} \left(\frac{ez}{2\nu} \right)^{-\nu}.
\end{eqnarray} 
\endnumparts 
Taking into account these formulas in the expression for $g_l(k)$ we obtain
\begin{equation}
\nu g_l(k) \approx  \frac{\nu^3 \chi^{-1-2\nu}}{\nu Q + 2 (kR)^2}.  
\end{equation}
Now we integrate over $k$ with factor $Q$ in Eq. (\ref{eq:T0})
\begin{equation}
Q \int_0^\infty \frac{dk}{k^2 + k_a^2} \frac{1}{\nu Q + 2 (kR)^2} = \frac{\pi }{2k_a } \frac{\sqrt{Q}}{\sqrt{Q}\nu + \sqrt{2\nu} q_a},
\end{equation}
and obtain series
\begin{equation}\label{eq:Q}
\sum_{l=1}^\infty \frac{\pi }{2k_a } \frac{\sqrt{Q}\nu^3 \chi^{-1-2\nu}}{\sqrt{Q}\nu + \sqrt{2\nu} q_a},
\end{equation}
in which we take limit of short distances, $\chi = 1+r \to 1$. If $Q \not = 0$ we obtain 
\begin{equation}
\sum_{l=1}^\infty \frac{\pi }{2k_a } \nu^2 \chi^{-1-2\nu} =  \frac{\pi }{8k_a } \frac{1- 2 \chi^2 + 9 \chi^4}{\chi^2 (\chi^2 - 1)^3} \approx \frac{\pi }{8k_a r^3}.
\end{equation}
Therefore, we arrive with expression 
\begin{equation}
E_0 = - \frac{\hbar \omega_a \alpha (0)}{8d^3},
\end{equation}
which is valid for $r = d/R \ll Q = \Omega R$. In the case of $Q = 0$ we obtain from Eq. (\ref{eq:Q}) that $E_0 = 0$. 

Therefore, for $Q >0$ and $r = d/R \ll Q$ we obtain free energy
\begin{equation}\label{eq:Fsd}
{\cal F}_a = - \frac{\alpha (0)}{4d^3} \left\{ \frac{\hbar \omega_a}{2} + \frac{\hbar \omega_a}{e^{\frac{\hbar \omega_a}{k_B T}} -1}\right\}.
\end{equation} 

This expression is in agreement with high and low temperature expansions. Indeed, taking the corresponding limits in (\ref{eq:Fsd}) we obtain
\numparts 
\begin{eqnarray}
{\cal F}_a &=&  -\frac{k_B T\alpha (0)}{4d^3},\ T\to\infty, \\
{\cal F}_a &=&  -\frac{\hbar \omega_a \alpha (0)}{8d^3},\ T\to 0,
\end{eqnarray}   
\endnumparts 
which coincides with Eqs. (\ref{eq:FTsd}) and (\ref{eq:FaLowT}). 

\section{The entropy}\label{sec:4}

Taking the minus derivative with respect to the temperature (with $k_B$) from the free energy given by Eqs. (\ref{eq:T2}) we obtain expression for entropy $S = S_1 + S_2$ ($a = T_\omega/T$):
\numparts 
\begin{eqnarray}
\fl S_1 = \frac{\alpha(0)}{R^3} \frac{e_1(a)}{(1+r)^2} \left(\frac{\pi a}{\sinh \pi a} \right)^2, \\
\fl S_2 = \frac{\alpha(0)}{R^3} \frac{1}{\pi (1+r)^2} \int_0^\infty \left\{t e'_2  + e_2 (1-2\pi a t \coth \pi a t)\right\} \ln\left|\frac{1+t}{1-t}\right| \left(\frac{\pi a}{\sinh \pi a} \right)^2 dt. 
\end{eqnarray}
\endnumparts 
\begin{figure}[ht]
\centerline{\includegraphics[width=7cm]{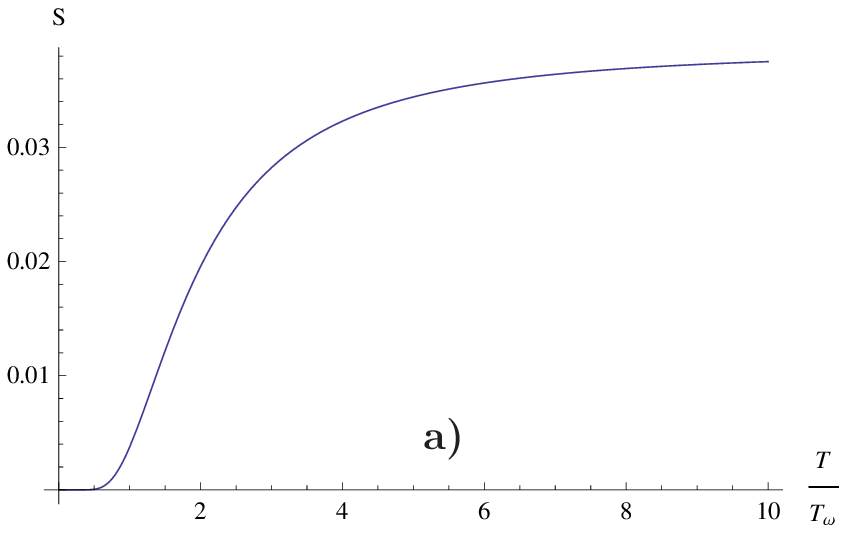}\includegraphics[width=7cm]{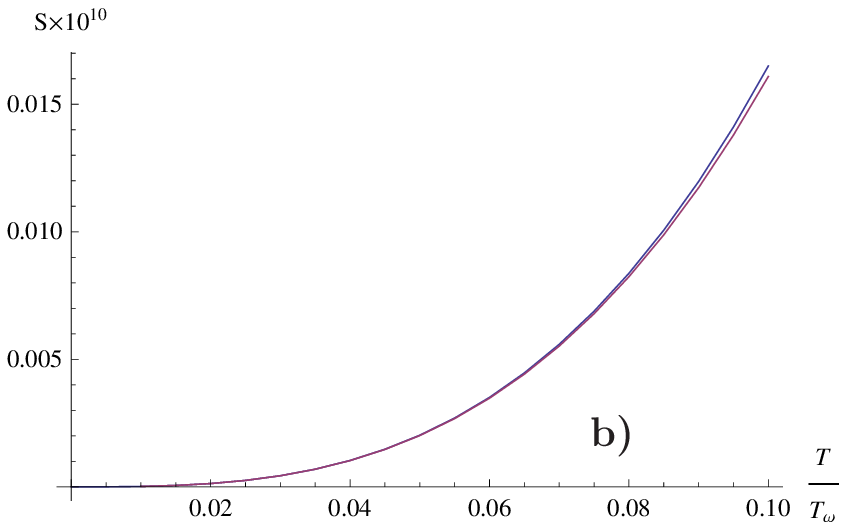}}
\caption{The plot of the entropy $S$ calculated for the hydrogen atom and the sphere with parameters of the fullerene $C_{60}$ and distance from fullerene $d=R/2\  (r=0.5)$. The entropy for large interval of temperature is shown in Fig. a). In Fig. b) we show the entropy for low temperature. The exact numerical calculation is upper curve and lower curve is approximate expression (\ref{eq:EntSmallT}).}\label{fig:ent-1}
\end{figure}
The entropy at low and high temperatures maybe found from Eqs. (\ref{eq:FaLowT-A}) and (\ref{eq:FHighT}) namely,
\numparts 
\begin{eqnarray}
\fl S = \frac{16\pi^3}{15} \frac{\alpha (0)}{(1+r)^6} \left(\frac{k_B T}{\hbar c} \right)^3 = \frac{2}{15} \frac{\alpha (0) k_a^3}{(1+r)^6} \left(\frac{T}{T_\omega} \right)^3 , \  T \ll T_R,T_\omega,T_{R+d}, \label{eq:Entr-1}\\
\fl S = \frac{\alpha (0)}{R^3}\frac{6 r^4+24 r^3+33 r^2+18 r+4}{2 r^3 (r+1)^4 (r+2)^3}, \ T \gg T_R,T_\omega,T_{R+d} . \label{eq:EntSmallT}
\end{eqnarray}
\endnumparts 
The typical behaviour of the entropy is shown in the Fig. \ref{fig:ent-1} for sphere with parameters of the fullerene $C_{60}$ and hydrogen atom with distance between them $d=R/2,\ (r=0.5)$. For this situation we have the following numbers: $R = 3.42 \mathring{A},\ Q = \Omega R = 4.94\cdot 10^{-2},\ \alpha (0) =0.667 \mathring{A}^3,\ q_a = \omega_a R/c = 0.0202$.   

From the Fig. \ref{fig:ent-1} we observe that the Casimir-Polder entropy tends to zero as third power of the temperature and it is zero at zero temperature according the Nernst heat theorem. It is positive for arbitrary temperatures. In Ref. \cite{Bezerra:2008:Ltaiatqr} was observed the  negative sign of the entropy for short separations, $r\to 0$, and low temperatures in the case of the atom with plate. We note that the low temperature dependence given by Eq. (\ref{eq:Entr-1}) is valid for $T\ll T_R = \hbar c/2\pi k_B R$. To compare with the case of atom near the flat boundary considered in Ref. \cite{Bezerra:2008:Ltaiatqr} we should take the limit $R\to \infty$ and in this case the relation $T\ll T_R = \hbar c/2\pi k_B R$ no longer be valid.   

\begin{figure}[ht]
\centerline{\includegraphics[width=7cm]{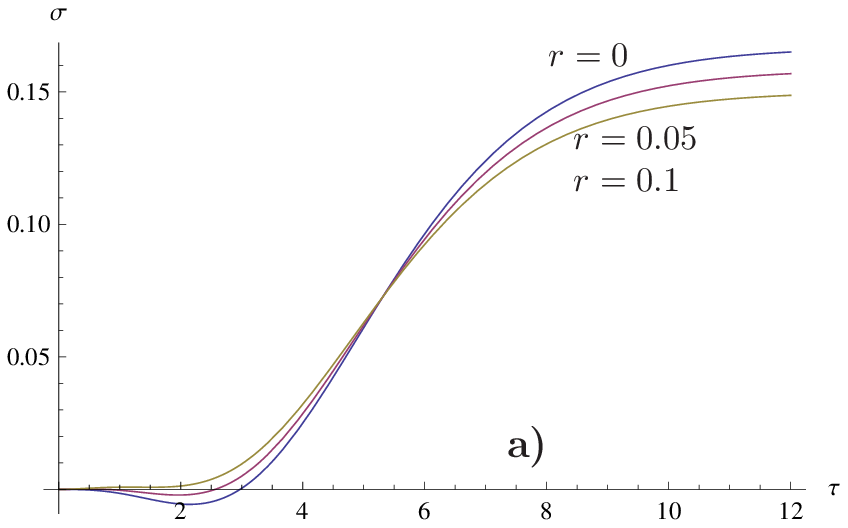}\includegraphics[width=7cm]{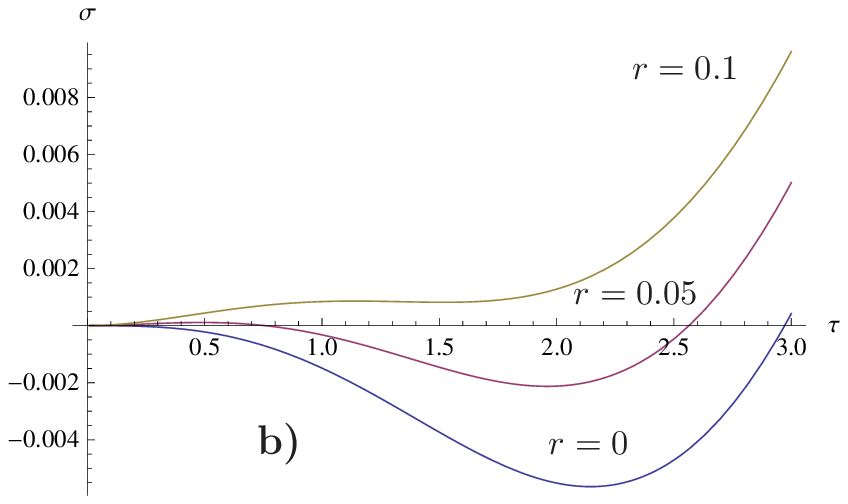}}
\caption{The plot of the function $\sigma$ calculated for different values of $r = d/R = 0,\ 0.05,\ 0.1$.  The plot a) shows the function for large domain of $\tau = 4\pi k_B dT/\hbar c$. In Fig. b) we show the function $\sigma$ close to the origin. Starting with $r = d/R = 0.085$ the entropy becomes completely positive.}\label{fig:ent-3}
\end{figure}

To show appearance the positive entropy for finite radius of sphere and negative sign of entropy for infinite radius of sphere we use expansion over $1/R$ obtained above (\ref{eq:FgenBR}). For the case of static polarizability, $\alpha (i\omega) = \alpha (0)$, we obtain from Eq. (\ref{eq:FgenBR})
\begin{equation}\label{eq:static}
{\cal F}^a = E_{CP} \left(\eta_0 + \frac{d}{R} \eta_1 \right),
\end{equation}   
where
\begin{eqnarray*}
\fl \eta_0 = \frac{\tau}{6} \left\{ 1 + \frac{2}{e^\tau - 1} + \frac{2\tau e^\tau}{(e^\tau - 1)^2} + \frac{\tau^2 e^\tau (e^\tau +1)}{(e^\tau -1)^3} \right\}, \\
\fl \eta_1 = -\frac{\tau}{6} \left\{ 1 + \frac{2}{e^\tau - 1} + \frac{2\tau e^\tau}{(e^\tau - 1)^2} + \frac{3\tau^2 e^\tau (e^\tau +1)}{2(e^\tau -1)^3} + \frac{\tau^3 e^\tau (e^{2\tau} + 4 e^\tau +1)}{2(e^\tau -1)^4}\right\}\nonumber\\
 + \frac{\tau^5}{16} \int_{\tau/2}^\infty \frac{dt}{t} \frac{\cosh t}{\sinh^5 t} + \frac{\tau^3(\tau^2-4)}{48} \int_{\tau/2}^\infty \frac{dt}{t} \frac{\cosh t}{\sinh^3 t},
\end{eqnarray*}
and $\tau = 2T/T_d = 4\pi k_B dT/\hbar c$. The function $\eta_0$ was found in Ref. \cite{Bezerra:2008:Ltaiatqr}, the function $\eta_1$ gives the correction due to finite radius of sphere. Now we use these expressions to calculate the entropy. Taking the minus derivative with respect to the $k_B T$ we obtain the following expression for entropy 
\begin{equation}
S = \frac{3\alpha (0)}{2d^3} \sigma,
\end{equation}
where $\sigma = \eta'_{0,\tau} + r \eta'_{1,\tau}$. The numerical calculations of the entropy for large $R$ by using formula (\ref{eq:static}) for different values of $r = d/R = 0,\ 0.05,\ 0.1$ are shown in the Fig. \ref{fig:ent-3}. Numerically we found that for $r = d/R > 0.085$ the entropy becomes completely positive. 

\section{Conclusion} \label{sec:5}
In above sections we considered in detail the thermal Casimir-Polder interaction of an atom with spherical conductive surface which models fullerene. For sphere we use plasma (hydrodynamic) model \cite{Barton:2004:Cesps,Barton:2005:CefpsIE}: the sphere is two-dimensional conductive surface with finite conductivity. All information about sphere is encoded in one parameter $\Omega = 4 \pi n e^2/m c^2$ with dimension of wave number. To obtain the free energy we adopt the Lifshitz approach in framework of which the spherical shell is situated inside the spherical cavity into the infinite dielectric media. Rarefying this media we obtain the free energy per unit atom. Renormalization procedure is simple -- we throw away all terms which survive in the limit $\Omega \to 0$ because this case corresponds to the case without sphere and the energy should be zero. The opposite case $\Omega \to \infty$ corresponds to the ideal case.  The general expression for the free energy is given by Eq. (\ref{eq:Fgen}) 
or Eq. (\ref{eq:T2}). 

We made analysis of this expression for different cases. In the limit of infinite radius, $R$, of the sphere with finite distance, $d$, between an atom and sphere we obtain the expected expression for the energy for the case of the atom close to the plate. We also found the next term of expansion $\sim d/R$. In the problem under consideration we have three different parameters with dimension of the temperature:
\begin{equation}
T_\omega = \frac{\hbar \omega_a}{2\pi k_B},\ T_R = \frac{\hbar c}{2\pi k_B R}, \ T_d = \frac{\hbar c}{2\pi k_B d}.
\end{equation}
Each temperature connects with specific scale namely, first connects with scale of atom, the second -- with scale of sphere and the third connects with distance between an atom and the sphere. For molecule of fullerene $C_{60}$ we have $T_\omega = 2.15\cdot 10^4 K,\ T_R = 1.06\cdot 10^6 K$. For low temperatures, $T \ll T_\omega,T_R,T_d$ we obtain (see Eq. (\ref{eq:FaLowT})) that the free energy has temperature correction proportional to the forth power of the temperature $\sim T^4$. The high temperature expansion, $T \gg T_\omega,T_R,T_d$ given by Eq. (\ref{eq:FHighT}) reveals the first power of the temperature.   
Nearby to the sphere the energy has the form given by Eq. (\ref{eq:Fsd}). It has the form of the sum the zero temperature term and the Plank form temperature correction. 

The entropy of this system shows interesting behaviour (see Fig. \ref{fig:ent-1}). The entropy is positive for any distance between atom and sphere of small radius. It tends to zero for low temperatures according to the Nernst heat theorem and it tends to constant for high temperatures. It has been shown in the Ref.  \cite{Bezerra:2008:Ltaiatqr} that the entropy is negative for short distances between an atom and the flat boundary. To reveal  the sign changing we found expansion of the entropy over $d/R$ up to the fist power of $d/R$. The analytical and numerical analysis shows that the entropy indeed has the region of the negative sign for large radii of the sphere and becomes completely positive beginning  with some definite value of the $d/R$ (see Fig. \ref{fig:ent-3}). 

As noted above in Introduction the hydrodynamical (plasma) approach has limited applications and the more realistic model to describe the interaction between an atom with fullerene is the Dirac model. It is based on the $2+1$ dimensional Dirac equation with very small parameter of the mass and the Fermi velocity instead of the velocity of light. The future aim of our consideration is the Casimir-Polder interaction in the framework of the Dirac model.          

\section*{Acknowledgements}
The author would like to thank M. Bordag for critical reading of the paper and V. Mostepanenko and G. Klimchitskaya for discussions. This work was supported by the Russian Foundation for Basic Research Grant No. 11-02-01162-a.

\section*{References}

\input{cptiop-bib}
\end{document}

%% file: cptiop-bib.tex
\providecommand{\newblock}{}